\documentclass[pra,aps,amsmath,amssymb,showkeys,showpacs]{revtex4}

\newcommand{\cald}{\mathcal{D}}
\newcommand{\id}{\mathrm{id}}
\newcommand{\hh}{\mathcal{H}}
\newcommand{\lnp}{\mathcal{L}}

\newcommand{\lsp}{\mathcal{L}_{+}}
\newcommand{\lpp}{\mathcal{L}_{++}}
\newcommand{\bro}{\boldsymbol{\rho}}
\newcommand{\km}{\mathsf{K}}

\newcommand{\vm}{\mathsf{V}}

\newcommand{\ax}{\mathsf{X}}
\newcommand{\ay}{\mathsf{Y}}

\newcommand{\bomg}{\boldsymbol{\omega}}

\newcommand{\D}{\mathsf{D}}
\newcommand{\pen}{\openone}

\newcommand{\rmr}{\mathrm{R}_{q}^{(s)}}
\newcommand{\rmm}{\mathrm{M}_{q}^{(s)}}

\newcommand{\Tr}{\mathrm{Tr}}
\newcommand{\vc}{\mathrm{vec}}

\begin{document}
\clearpage
\preprint{}

\title{Unified-entropy trade-off relations for a single quantum channel}
\author{Alexey E. Rastegin}
\affiliation{Department of Theoretical Physics, Irkutsk State University,
Gagarin Bv. 20, Irkutsk 664003, Russia}

\begin{abstract}
Many important properties of quantum channels are quantified by
means of entropic functionals. Characteristics of such a kind are
closely related to different representations of a quantum channel.
In the Jamio{\l}kowski--Choi representation, the given quantum
channel is described by the so-called dynamical matrix. Entropies
of the rescaled dynamical matrix known as map entropies describe a
degree of introduced decoherence. Within the so-called natural
representation, the quantum channel is formally posed by another
matrix obtained as reshuffling of the dynamical matrix. The
corresponding entropies characterize an amount, with which the
receiver {\it a priori} knows the channel output. As was
previously shown, the map and receiver entropies are mutually
complementary characteristics. Indeed, there exists a non-trivial
lower bound on their sum. First, we extend the concept of receiver
entropy to the family of unified entropies. Developing the
previous results, we further derive non-trivial lower bounds on
the sum of the map and receiver $(q,s)$-entropies. The derivation
is essentially based on some inequalities with the Schatten norms
and anti-norms.
\end{abstract}



\maketitle

\pagenumbering{arabic}
\setcounter{page}{1}

\section{Introduction}\label{sec1}

One of fundamental distinctions of the quantum world from the
classical world is expressed by the uncertainty principle
\cite{heisenberg}. We mention here only several developments
concerning this issue. More extensive discussion with the
corresponding references can be found in the reviews
\cite{lahti,ww10,brud11}. Recent studies of entropic relations are
often connected with advances in quantum information science. It
treats quantum states and effects as informational resources
\cite{nielsen,watrous1}. Due to the influential works
\cite{deutsch,maass}, entropic functionals are now typically used
in formulating the uncertainty principle. Entropic relations are
essential in analyzing the security of quantum cryptographic
schemes \cite{dmg07,ngbw12}. Reformulations in the presence of
quantum memory are discussed in \cite{BCCRR10,ccyz12,fan12}. With
respect to quantum operations, an entropic approach has been
developed in \cite{rprz12}. In expressing entropic relations,
various kinds of entropic functionals have found to be useful. Such
entropic bounds were utilized in studying several specific topics
such as quantifying number-phase uncertainties
\cite{rast105,rast12num}, incompatibilities of anti-commuting
observables \cite{ww08} and reformulations in quasi-Hermitian
models \cite{rast12quasi}. The R\'{e}nyi \cite{renyi} and Tsallis'
entropies \cite{tsallis} are both especially important. They are
particular cases of unified $(q,s)$-entropies introduced in
\cite{hey06} and further studied in \cite{rastjst}. Formally,
entropic uncertainty relations are posed as lower bounds on the
sum of two or more entropies. Each of these entropies characterize
an uncertainty in measurement statistics with respect to the
corresponding basis.

In the paper \cite{rprz12}, the authors derived a lower bound on
the sum of two entropies, corresponding to complementary
characteristics of the same quantum operation. Inequalities of
such a kind are usually referred to as trade-off relations. The
formalism of quantum operations provides a general tool for
description of open quantum systems. Deterministic protocols are
represented by trace-preserving operations also known as quantum
channels. In quantum information theory, several channel
representations are significant. The Jamio{\l}kowski--Choi and
natural representations are both utilized in entropic
characterizing given quantum channel. The entropy exchange
\cite{glind91,bsch96} and the map entropy \cite{zb04} are
respectively used to describe entanglement transmission and
decoherence induced by a quantum channel. A family of map
entropies is defined in terms of the dynamical matrix. The writers
of \cite{rprz12} have defined the receiver entropies related to
the natural representation of a channel. They further showed that
the sum of the map and receiver entropies is bounded from below.
This bound is written in terms of system dimensionality and
compares two complementary characteristics of a single quantum
channel. Thus, it differs in character from uncertainty relations
posed in \cite{rast104} for extremal unravelings of two channels.
The writers of \cite{rprz12} utilized R\'{e}nyi's entropic
functionals. Other entropic functionals have found use in studying
various quantum properties
\cite{raggio,majer03,kim11,rastfn,rastctp}.

The aim of the present work is to formulate trade-off relations
for a single quantum channel in terms of the map and receiver
$(q,s)$-entropies. The paper is organized as follows. In section
\ref{sec2}, the required material is reviewed. First, definitions
of the Schatten norms and anti-norms are discussed. Second, we
briefly recall the concept of map entropies studied in
\cite{zb04,rfz10,rzf11,rast11a}. Then, the family of receiver
entropies are extended in terms of quantum $(q,s)$-entropies.
Section \ref{sec3} is devoted to formulating the main results.
Before their derivation, we collect several auxiliary statements.
In particular, some relations with the Schatten norms and
anti-norms are obtained. These inequalities could be used in other
questions, in which we deals with norms and anti-norms of matrices
related by reordering the same entries. Non-trivial lower bounds
on the sum of the map and receiver $(q,s)$-entropies are obtained
for all considered values of the parameters. Here, we establish
lower bounds for arbitrary quantum channels as well as stronger
bounds for unital channels. In section \ref{sec4}, we conclude the
paper with a summary of results.

\section{Definitions and notation}\label{sec2}

In this section, we introduce terms and conventions that will be
used through the text. First, required material on operator norms
and anti-norms is presented. Second, we discuss the map
$(q,s)$-entropy. Then the notion of the receiver $(q,s)$-entropy
is introduced.

\subsection{Operators, norms and anti-norm}\label{sc21}

Let $\lnp(\hh)$ be the space of linear operators on
$d$-dimensional Hilbert space $\hh$. By $\lsp(\hh)$ and
$\lpp(\hh)$, we respectively denote the set of positive
semidefinite operators on $\hh$ and the set of strictly positive
ones. A density operator $\bro\in\lsp(\hh)$ has unit trace, i.e.
$\Tr(\bro)=1$. For $\ax,\ay\in{\mathcal{L}}(\hh)$, we define the
Hilbert--Schmidt inner product by \cite{watrous1}
\begin{equation}
\langle\ax{\,},\ay\rangle_{\rm{hs}}:=\Tr(\ax^{\dagger}\ay)
\ . \label{hsdef}
\end{equation}
The Schatten norms, which form an important class of unitarily
invariant norms, are defined in terms of singular values. The
singular values $\sigma_{j}(\ax)$ of operator $\ax$ are put as
eigenvalues of $|\ax|=\sqrt{\ax^{\dagger}\ax}\in\lsp(\hh)$. For
$q\geq1$, the Schatten $q$-norm of
$\ax\in{\mathcal{L}}({\mathcal{H}})$ is then defined by
\cite{watrous1,bhatia97}
\begin{equation}
\|{\mathsf{\ax}}\|_{q}:=
\left(\sum\nolimits_{j=1}^{d}\sigma_{j}({\mathsf{\ax}})^q\right)^{1/q}
=\Tr\!\left(\bigl(\ax^{\dagger}\ax\bigr)^{q/2}\right)^{1/q}
{\,}. \label{shndf}
\end{equation}
Here, the singular values should be inserted according to their
multiplicities. The family (\ref{shndf}) gives the trace norm
$\|\ax\|_{1}$ for $q=1$, the Frobenius norm $\|\ax\|_{2}$ for
$q=2$, and the spectral norm
$\|\ax\|_{\infty}=\max\bigl\{\sigma_j(\ax):{\>}1\leq{j}\leq{d}\bigr\}$
for $q=\infty$. Note that the Frobenius norm can be rewritten as
\begin{equation}
\|\ax\|_{2}=\langle\ax{\,},\ax\rangle_{\rm{hs}}^{1/2}
\ . \label{fvhs}
\end{equation}
For this reason, it is often called the Hilbert--Schmidt norm.
This norm induces the Hilbert--Schmidt distance, which is very
useful to analyze a geometry of quantum states \cite{dunk11}. The
formula (\ref{fvhs}) shows that the squared $2$-norm is expressed
as the sum of squared modulus of all the matrix entries. In the
following, this fact will be essential.

In the papers \cite{bh10a,bh10b}, Bourin and Hiai examined
symmetric anti-norms of positive operators. They form a class of
functionals containing the right-hand side of (\ref{shndf}) for
$q\in(0;1)$ and, with strictly positive matrices, for $q<0$. For
arbitrary $\ax\in\lsp(\hh)$, we consider a functional
$\ax\mapsto\|\ax\|_{!}$, taking values on $[0;\infty)$. If this
functional enjoys the homogeneity, the symmetry
$\|\ax\|_{!}=\|\vm\ax\vm^{\dagger}\|_{!}$ for all unitary $\vm$,
and the superadditivity
\begin{equation}
\|\ax+\ay\|_{!}\geq\|\ax\|_{!}+\|\ay\|_{!}
\ , \label{sups}
\end{equation}
it is said to be a symmetric anti-norm \cite{bh10a}. Let
$\lambda_{j}({\mathsf{\ax}})$ denote $j$-th eigenvalue of operator
$\ax$. For $\ax\in\lpp(\hh)$ and $q\leq1\neq0$, we put the
Schatten $q$-anti-norm as
\begin{equation}
\|{\mathsf{\ax}}\|_{q}=\left(\sum\nolimits_{j=1}^{d}\lambda_{j}({\mathsf{\ax}})^q\right)^{1/q}
{\,}. \label{shnaf}
\end{equation}
Using given anti-norm $\|\ax\|_{!}$, for any $q\in(0;1)$ we can
build another anti-norm (see proposition 3.7 of \cite{bh10a})
\begin{equation}
\ax\mapsto\|\ax^{q}\|_{!}^{1/q}
\ . \label{afunex}
\end{equation}
Note that the trace norm is actually an anti-norm on positive
matrices. Indeed, the superadditivity inequality is satisfied here
with equality. Combining this with (\ref{afunex}), the right-hand
side of (\ref{shnaf}) is an anti-norm on positive operators for
$q\in(0;1)$. With $\ax\in\lpp(\hh)$ and arbitrary non-zero
$q\leq1$, the inequality (\ref{sups}) for the Schatten
$q$-anti-norm was proved in \cite{rastcejp}. Note that the
restriction to positive matrices is essential in the sense of
superadditivity condition (\ref{sups}). Following
\cite{bh10a,bh10b}, the Schatten $q$-anti-norm of $\ax$ will also
be denoted as $\|\ax\|_{q}$. Some relations for anti-norms and
their applications to entropic functionals are examined in
\cite{rastjst12}. We will further assume that use of anti-norms
tacitly implies the corresponding restriction on matrices. For all
$0<p<q$, we have
\begin{equation}
\|\ax\|_{q}\leq\|\ax\|_{p}
\ . \label{npqr}
\end{equation}
This relation is actually no more than theorem 19 of \cite{hardy}.
In the following, we will use the Schatten $q$-norms for
$q\in[1;\infty]$ and the Schatten $q$-anti-norms of positive
operators for $q\in(0;1)$.

\subsection{Two channel representations and the corresponding $(q,s)$-entropies}\label{sc22}

By $\hh$, we mean $d$-dimensional space of the principal quantum
system $Q$. Consider a linear map
$\Phi{\>}:\lnp(\hh)\rightarrow\lnp(\hh)$, which also satisfies the
condition of complete positivity. Let $\id^{R}$ be the identity
map on $\lnp(\hh_{R})$, where the space $\hh_{R}$ is assigned to
auxiliary reference system. The complete positivity implies that
$\Phi\otimes\id^R$ transforms a positive operator into a positive
operator again for all extensions. When the identity operator is
mapped into itself, i.e. $\Phi(\pen)=\pen$, the map is unital.
Unital quantum channels are often called bistochastic
\cite{bengtsson}. There are many interesting classes of quantum
channels such as unistochastic channels and degradable channels.
The former implies that the principal system unitarily interacts
with the $d$-dimensional environment, which was initially
completely mixed \cite{bengtsson}. The notion of degradable
channels is related to studying capacities for various processing
scenarios \cite{des05}.

Linear maps can formally be described in several ways. We will use
two representations, when the map is represented by a single
matrix of size $d^{2}\times{d}^{2}$. The family of map entropies
is introduced within the Jamio{\l}kowski--Choi representation
\cite{jam72,choi75}, for which we take $\hh_{R}=\hh$. To fixed
orthonormal basis $\{|\nu\rangle\}$ in $\hh$, we assign the
normalized pure state
\begin{equation}
|\phi_{+}\rangle:=\frac{1}{\sqrt{d}}{\,}\sum_{\nu=1}^{d}
{|\nu\rangle\otimes|\nu\rangle}
\ . \label{phip}
\end{equation}
One defines the operator
$\bomg(\Phi):=\Phi\otimes\id{\,}\bigl(|\phi_{+}\rangle\langle\phi_{+}|\bigr)$,
acting on the space $\hh\otimes\hh$. The dynamical matrix is then
written as $\D(\Phi)=d{\,}\bomg(\Phi)$ \cite{zb04}. In shortened
notation, we write the dynamical matrix and the rescaled one as
$\D_{\Phi}$ and $\bomg_{\Phi}$, respectively. For all
$\ax\in{\mathcal{L}}(\hh)$, the action of $\Phi$ can be expressed
as \cite{watrous1}
\begin{equation}
\Phi(\ax)=\Tr_{R}\left(\D_{\Phi}(\pen\otimes\ax^{T})\right)
{\>}, \label{chjis}
\end{equation}
where $\ax^{T}$ denotes the transpose operator to $\ax$. The map
$\Phi$ is completely positive, if and only if the matrix
$\D(\Phi)$ is positive. Further, the preservation of the trace is
expressed as $\Tr_{Q}(\D_{\Phi})=\pen$, whence $\Tr(\D_{\Phi})=d$
and $\Tr(\bomg_{\Phi})=1$.

The concept of entropy is one of fundamentals in both statistical
physics and information theory \cite{wehrl,petz08}. We focus an
attention on the notions of the map and receiver entropies. The
former is put through the Jamio{\l}kowski--Choi representation.
For any quantum channel, the dynamical matrix is positive one with
trace $d$. For $q>0\neq1$ and $s\neq0$, the map $(q,s)$-entropy of
quantum channel $\Phi$ is defined as
\begin{equation}
\rmm(\Phi):=\frac{1}{(1-q){\,}s}{\,}
\left(\frac{\|\D_{\Phi}\|_{q}^{qs}}{d^{qs}}-1\right)
=\frac{1}{(1-q){\,}s}{\,}
\left\{\left(d^{-q}\sum\nolimits_{j=1}^{d^{2}}{\,}\lambda_{j}(\D_{\Phi})^{q}\right)^{\!{s}}-1\right\}
{\>}. \label{mapqs}
\end{equation}
In the limit $s\to0$, one leads to the map R\'{e}nyi entropy
defined as
\begin{equation}
\mathrm{M}_{q}^{(0)}(\Phi):=
\frac{q}{1-q}{\>}\Bigl(\ln\|\D_{\Phi}\|_{q}-\ln{d}\Bigr)=
\frac{1}{1-q}{\,}\left\{\ln\!\left(\sum\nolimits_{j=1}^{d^{2}}\lambda_{j}(\D_{\Phi})^q\right)-q\ln{d}\right\}
{\>}. \label{mapqs0}
\end{equation}
Taking $s=1$, we have the case of Tsallis entropy
\begin{equation}
\mathrm{M}_{q}^{(1)}(\Phi)=\frac{d^{-q}{\,}\|\D_{\Phi}\|_{q}^{q}-1}{1-q}=
-\sum_{j=1}^{d^{2}}\left(\frac{\lambda_{j}(\D_{\Phi})}{d}\right)^{\!q}\ln_{q}\!\left(\frac{\lambda_{j}(\D_{\Phi})}{d}\right)
{\>}, \label{tmap}
\end{equation}
where the $q$-logarithm $\ln_{q}(x)=\bigl(x^{1-q}-1\bigr)/(1-q)$
is defined for $q>0\neq1$ and $x>0$. When $q\to1$, we obtain the
usual logarithm and the map von Neumann entropy. For brevity, we
will usually omit the summation range, when it is clear from the
context. The expression (\ref{mapqs}) is a two-parametric
extension of the standard map entropy introduced in \cite{zb04}.
Note that the map entropy coincides with the entropy exchange
calculated for given channel with the completely mixed input
\cite{rfz10,wfz08}. Properties of the map $(q,s)$-entropies are
examined in \cite{rast11a,rastcejp}. Therein, the map
$(q,s)$-entropy was introduced in terms of the rescaled matrix
$\bomg_{\Phi}$. In the present consideration, however, the
definitions (\ref{mapqs}) and (\ref{mapqs0}) are more convenient.

To consider the natural representation, we will use a variant of
operator-vector correspondence
$\lnp(\hh)\rightarrow\hh\otimes\hh$. With respect to the standard
basis $\{|\nu\rangle\}$, one defines the mapping \cite{watrous1}
\begin{equation}
\big|\vc(|\mu\rangle\langle\nu|)\bigr\rangle=|\mu\rangle\otimes|\nu\rangle
\ . \label{vcdf}
\end{equation}
By linearity, this mapping is determined for arbitrary input
$\ax\in\lnp(\hh)$. It is an isometry in the sense that
$\langle\ax{\,},\ay\rangle_{\mathrm{hs}}=\bigl\langle\vc(\ax)\big|\vc(\ay)\bigr\rangle$
\cite{watrous1}. We now define the matrix
$\km(\Phi)\in\lnp(\hh\otimes\hh)$ such that, for all
$\ax\in\lnp(\hh)$,
\begin{equation}
\big|\vc\bigl(\Phi(\ax)\bigr)\bigr\rangle=\km(\Phi)\big|\vc(\ax)\bigr\rangle
\ . \label{kmdf}
\end{equation}
Following \cite{rprz12,wfz08}, this matrix will be referred to as
the superoperator matrix. Some properties of superoperator
matrices are indicated in section 5.2.1 of \cite{watrous1}. Note
that the superoperator matrix is not generally Hermitian. In
characterizations of quantum channels, the natural representation
is not so frequently used as other representations
\cite{watrous1}. Nevertheless, the authors of \cite{rprz12} have
shown that entropic characteristics of $\km_{\Phi}$ are
demonstrative as well. Namely, the family of receiver entropies is
defined in terms of the superoperator matrix \cite{rprz12}. In
line with the right-hand sides of (\ref{mapqs}) and
(\ref{mapqs0}), we define the receiver $(q,s)$-entropy by
\begin{align}
\rmr(\Phi)&:=\frac{1}{(1-q){\,}s}{\,}
\left\{\left(\|\km_{\Phi}\|_{1}^{-q}\sum\nolimits_{j=1}^{d^{2}}{\,}\sigma_{j}(\km_{\Phi})^{q}\right)^{\!{s}}-1\right\}
\qquad (s\neq0) \ , \label{repqs}\\
\mathrm{R}_{q}^{(0)}(\Phi)&:=
\frac{1}{1-q}{\,}\left\{\ln\!\left(\sum\nolimits_{j=1}^{d^{2}}\sigma_{j}(\km_{\Phi})^{q}\right)-q\ln\|\km_{\Phi}\|_{1}\right\}
{\>}, \label{repqs0}
\end{align}
where $\|\km_{\Phi}\|_{1}=\sum_{j}\sigma_{j}(\km_{\Phi})$. This
definition can be rewritten in terms of the Schatten $q$-norm of
$\km_{\Phi}$ for $q\geq1$ and $q$-anti-norm of positive
$|\km_{\Phi}|$ for $q\in(0;1)$. The right-hand side of
(\ref{repqs0}), i.e. R\'{e}nyi's variety, was previously
introduced in \cite{rprz12}. It characterizes an amount, with
which the receiver {\it a priori} knows the channel output. For
our purposes, the following relation between the above matrices is
essential. The standard basis in $\hh\otimes\hh$ is formed by the
vectors $|\mu\nu\rangle\equiv|\mu\rangle\otimes|\nu\rangle$. For
arbitrary vectors of this basis, we have
\begin{equation}
\langle\alpha\beta|\km_{\Phi}|\mu\nu\rangle=
\langle\alpha\mu|\D_{\Phi}|\beta\nu\rangle
\ . \label{kmdm}
\end{equation}
In other words, the superoperator matrix is obtained from the
dynamical one by the reshuffling operation \cite{rprz12}. It is
for this reason that the matrices $\D_{\Phi}$ and $\km_{\Phi}$
instead of rescaled ones are to be used in our definitions. Matrix
operations of such a kind and their essential role in quantum
information are reviewed in \cite{jam11}.  The map end receiver
entropies provide two mutually complementary characteristics of
given quantum channel. This issue is considered in the next
section.

\section{Trade-off relations in terms of the $(q,s)$-entropies}\label{sec3}

In this section, we formulate trade-off relations for a single
quantum channel in terms of the map and receiver
$(q,s)$-entropies. To avoid too long derivation, we first present
several auxiliary results. In particular, some relations between
Schatten norms and anti-norms are given. Then, lower bounds on the
$(q,s)$-entropic sum are derived for all values of the parameters.

\subsection{Preliminaries}\label{sc31}

As was already mentioned, the dynamical and superoperator matrices
are $d^{2}\times{d}^{2}$-matrices related by the reshuffling
operation. Since the operation does not change the matrix entries,
these matrices have the same Schatten $2$-norm \cite{rprz12}.
Hence, relations between $2$-norm and other norms or anti-norms
are essential for our purposes.

\newtheorem{lem1}{Proposition}
\begin{lem1}\label{lan1}
For all $\ax\in\lnp(\hh)$ and $q\geq1$, the Schatten $q$-norm
satisfies
\begin{eqnarray}
\|\ax\|_{q}^{q}&\leq&\|\ax\|_{2}^{2(q-1)}{\,}\|\ax\|_{1}^{2-q}
\qquad (1\leq{q}\leq2)
\ , \label{lq1}\\
\|\ax\|_{q}^{q}&\geq&
\|\ax\|_{2}^{2(q-1)}{\,}\|\ax\|_{1}^{2-q}
\qquad (2\leq{q}<\infty)
\ . \label{lq2}
\end{eqnarray}
For $\ax\in\lsp(\hh)$ and $0<q<1$, the Schatten $q$-anti-norm
satisfies (\ref{lq2}). When $\ax\in\lpp(\hh)$, the relation
(\ref{lq2}) holds for $q<0$ as well.
\end{lem1}

{\bf Proof.} Let us introduce positive numbers
$x_{j}=\sigma_{j}(\ax){\,}\|\ax\|_{1}^{-1}$, which obey
$\sum_{j}x_{j}=1$. Assuming $q\geq1$, we apply Jensen's inequality
to the function $x\mapsto{x}^{q-1}$. As this function is concave
for $q\in[1;2]$ and convex for $q\in[2;\infty)$, we have
\begin{equation}
\sum\nolimits_{j} x_{j}^{q}=\sum\nolimits_{j} x_{j}{\,}x_{j}^{q-1}
\left\{
\begin{array}{cc}
\leq, & 1\leq{q}\leq2 \\
\geq, & 2\leq{q}<\infty
\end{array}
\right\}
\left(\sum\nolimits_{j} x_{j}^{2}\right)^{q-1}
{\,}. \label{jen1}
\end{equation}
The left-hand side of (\ref{jen1}) is equal to
$\|\ax\|_{q}^{q}{\,}\|\ax\|_{1}^{-q}$, the right-hand side of
(\ref{jen1}) is equal to
$\Bigl(\|\ax\|_{2}^{2}{\,}\|\ax\|_{1}^{-2}\Bigr)^{q-1}$. After
substituting these expressions, we respectively obtain (\ref{lq1})
and (\ref{lq2}). Restricting a consideration to strictly positive
$\ax$, we take $x_{j}=\lambda_{j}(\ax){\,}\|\ax\|_{1}^{-1}>0$.
Since the function $x\mapsto{x}^{q-1}$ is convex for $q<1$, the
above relation provides the claim (\ref{lq2}) for the Schatten
$q$-anti-norm. $\blacksquare$

We above mentioned that $\|\D_{\Phi}\|=\Tr(\D_{\Phi})=d$. Deriving
entropic relations, we will also use an upper bound on
$\|\km_{\Phi}\|_{\infty}$. The statement of theorem 1 in
\cite{rprz12} says that
\begin{equation}
\|\km_{\Phi}\|_{\infty}\leq{d}^{1/2}{\,}\|\Phi(\bro_{*})\|_{\infty}^{1/2}
\ , \label{upkp}
\end{equation}
where $\bro_{*}=\pen/d$ is the completely mixed state. Clearly,
one has
$\|\Phi(\bro_{*})\|_{\infty}\leq\|\Phi(\bro_{*})\|_{1}=\Tr\bigl(\Phi(\bro_{*})\bigr)=1$
for any quantum channel, whence
$\|\km_{\Phi}\|_{\infty}\leq{d}^{1/2}$. For unital channels, we
merely obtain
$\|\Phi(\bro_{*})\|_{\infty}=\|\bro_{*}\|_{\infty}=1/d$ and
$\|\km_{\Phi}\|_{\infty}\leq1$. It follows from lemma 3 of
\cite{rprz12} that, for arbitrary $\ax\in\lnp(\hh)$,
\begin{equation}
\|\ax\|_{2}\leq\|\ax\|_{\infty}^{1/2}{\,}\|\ax\|_{1}^{1/2}
\ . \label{21in}
\end{equation}
Combining the above facts with the equality
$\|\D_{\Phi}\|_{2}=\|\km_{\Phi}\|_{2}$, we then write
\begin{equation}
\frac{\|\D_{\Phi}\|_{1}}{\|\D_{\Phi}\|_{2}}{\>}\frac{\|\km_{\Phi}\|_{1}}{\|\km_{\Phi}\|_{2}}
=\frac{\|\D_{\Phi}\|_{1}{\,}\|\km_{\Phi}\|_{1}}{\|\km_{\Phi}\|_{2}^{2}}
\geq\frac{\|\D_{\Phi}\|_{1}}{\|\km_{\Phi}\|_{\infty}}
\geq
\left\{
\begin{array}{ll}
d^{1/2}{\>}, & {\mathrm{for{\ }all{\ }channels}}{\>}, \\
d{\>}, & {\mathrm{for{\ }unital{\ }ones}}{\>}.
\end{array}
\right.
\label{cbn0}
\end{equation}
Thus, the left-hand side of (\ref{cbn0}) is bounded from below. To
get trade-off relations from bounds on norms and anti-norms, some
analytical results are required as well. Let $a$ and $b$ be
strictly positive numbers such that $a<1$ and $1<b$. We consider
the following two-dimensional domains:
\begin{align}
\cald_{a}&:=\bigl\{(x,y):{\>}0\leq{x}\leq1,{\>}0\leq{y}\leq1,{\>}xy\leq{a}\bigr\}
\ , \label{dadf}\\
\cald_{b}&:=\bigl\{(x,y):{\>}1\leq{x}<\infty,{\>}1\leq{y}<\infty,{\>}b\leq{x}y\bigr\}
\ . \label{dbdf}
\end{align}
The minimum of the function $(x,y)\mapsto2-x-y$ in the domain
(\ref{dadf}) is equal to
\begin{equation}
{\min}\bigl\{2-x-y:{\>}(x,y)\in\cald_{a}\bigr\}=1-a
\ . \label{gmin}
\end{equation}
Indeed, this function decreases with each of $x$ and $y$. Hence,
the minimum is reached on the part of the curve $xy=a$ between the
points $(a,1)$ and $(1,a)$. As the function $x\mapsto2-x-a/x$ is
concave and takes the same value (\ref{gmin}) at the end points,
this value is actually the minimum. Further, the minimum of
$(x,y)\mapsto{x}+y-2$ in the domain (\ref{dbdf}) is equal to
\begin{equation}
{\min}\bigl\{x+y-2:{\>}(x,y)\in\cald_{b}\bigr\}=2\bigl(\sqrt{b}-1\bigr)
\ . \label{fmin}
\end{equation}
Since this function increases with each of $x$ and $y$, the
minimum is reached on the part of the curve $xy=b$ between $(b,1)$
and $(1,b)$. Then the claim is merely obtained by applying the
ordinary inequality between the arithmetic and geometric mean. The
results (\ref{gmin}) and (\ref{fmin}) will be used in obtaining
desired unified-entropy relations.

\subsection{Lower bounds on the sum of map and receiver entropies}\label{sc32}

Using the above preliminaries, we now derive lower bounds on the
sum of map and receiver $(q,s)$-entropies for all values of the
parameters.

\newtheorem{thm1}[lem1]{Proposition}
\begin{thm1}\label{tam1}
For $q>0$ and all real $s$, the sum of the map and receiver
$(q,s)$-entropies is bounded from below. For
$q\in(0;1)\cup(1;\infty)$ and $s\neq0$, the lower bound is written
as
\begin{align}
\rmm(\Phi)+\rmr(\Phi)&\geq\frac{\gamma}{s}{\>}
\ln_{q}\bigl(d^{s\varkappa/\gamma}\bigr)
\qquad{for{\ }all{\ }channels}
{\>}, \label{mnrs1}\\
\rmm(\Phi)+\rmr(\Phi)&\geq\frac{\gamma}{s}{\>}
\ln_{q}\bigl(d^{2s\varkappa/\gamma}\bigr)
\qquad{for{\ }unital{\ }ones}
{\>}. \label{mnrs2}
\end{align}
Here, the parameters $\gamma$ and $\varkappa$ are defined as
\begin{equation}
\gamma:=
\left\{
\begin{array}{ll}
1{\>}, & (1-q){\,}s<0{\>}, \\
2{\>}, & (1-q){\,}s>0{\>},
\end{array}
\right.
\qquad
\varkappa:=
\left\{
\begin{array}{ll}
1{\>}, & q\in(0;2]{\>}, \\
q\bigl(2(q-1)\bigr)^{-1}{\>}, & q\in[2;\infty){\>}.
\end{array}
\right.
\label{gamdf}
\end{equation}
\end{thm1}

{\bf Proof.} In the first place, we consider the case $q\in(0;1)$.
The sum of the map and receiver $(q,s)$-entropies is represented
as the function
\begin{equation}
f(x,y):=\frac{x+y-2}{(1-q){\,}s}
\ , \qquad x:=\frac{\|\D_{\Phi}\|_{q}^{qs}}{\|\D_{\Phi}\|_{1}^{qs}}
\ , \qquad y:=\frac{\|\ay_{\Phi}\|_{q}^{qs}}{\|\ay_{\Phi}\|_{1}^{qs}}
\ , \label{fudf}
\end{equation}
where $\ay_{\Phi}=|\km_{\Phi}|\in\lsp(\hh\otimes\hh)$. Suppose
also that $s>0$, then $x\geq1$ and $y\geq1$. Due to (\ref{lq2})
and (\ref{cbn0}), the anti-norms of positive matrices $\D_{\Phi}$
and $\ay_{\Phi}$ satisfy
\begin{equation}
\left(\frac{\|\D_{\Phi}\|_{q}}{\|\D_{\Phi}\|_{1}}{\>}\frac{\|\ay_{\Phi}\|_{q}}{\|\ay_{\Phi}\|_{1}}\right)^{\!q}\geq
\left(\frac{\|\D_{\Phi}\|_{1}}{\|\D_{\Phi}\|_{2}}{\>}\frac{\|\ay_{\Phi}\|_{1}}{\|\ay_{\Phi}\|_{2}}\right)^{\!2(1-q)}
\geq
\left\{
\begin{array}{ll}
d^{1-q}{\>}, & {\mathrm{for{\ }all{\ }channels}}{\>}, \\
d^{2(1-q)}{\>}, & {\mathrm{for{\ }unital{\ }ones}}{\>}.
\end{array}
\right.
\label{nrm0p}
\end{equation}
Therefore, the function $f(x,y)$ should be minimized in the domain
(\ref{dbdf}) with $b=d^{s(1-q)}$ for all quantum channels and
$b=d^{2s(1-q)}$ for unital channels. Using (\ref{fmin}), we obtain
the lower bound
\begin{equation}
\min\bigl\{f(x,y):{\>}(x,y)\in\cald_{b}\bigr\}=\frac{2}{s}
\left\{
\begin{array}{ll}
\ln_{q}\bigl(d^{s/2}\bigr){\>}, & {\mathrm{for{\ }all{\ }channels}}{\>}, \\
\ln_{q}\bigl(d^{s}\bigr){\>}, & {\mathrm{for{\ }unital{\ }ones}}{\>}.
\end{array}
\right.
\label{bun0p}
\end{equation}
Since $\gamma=2$ for $q\in(0;1)$ and $s>0$, the bound
(\ref{bun0p}) respectively concurs with (\ref{mnrs1}) and
(\ref{mnrs2}).

Defining $x$ and $y$ by (\ref{fudf}), we have $x\leq1$ and
$y\leq1$ in the case $q\in(0;1)$ and $s<0$. Since $(1-q){\,}s<0$,
the sum of the map and receiver $(q,s)$-entropies is rewritten as
the function
\begin{equation}
g(x,y):=\frac{2-x-y}{\bigl|(1-q){\,}s\bigr|}
\ . \label{gudf}
\end{equation}
Raising (\ref{nrm0p}) to the power $s<0$, we have arrived at the
following task. The function $g(x,y)$ should be minimized in the
domain (\ref{dadf}) with $a=d^{s(1-q)}$ for all quantum channels
and $a=d^{2s(1-q)}$ for unital channels.  Using (\ref{gmin}), we
obtain the lower bound
\begin{equation}
\min\bigl\{g(x,y):{\>}(x,y)\in\cald_{a}\bigr\}=\frac{1}{s}
\left\{
\begin{array}{ll}
\ln_{q}\bigl(d^{s}\bigr){\>}, & {\mathrm{for{\ }all{\ }channels}}{\>}, \\
\ln_{q}\bigl(d^{2s}\bigr){\>}, & {\mathrm{for{\ }unital{\ }ones}}{\>}.
\end{array}
\right.
\label{bun0m}
\end{equation}
Since $\gamma=1$ for $q\in(0;1)$ and $s<0$, the bound
(\ref{bun0m}) concurs with (\ref{mnrs1}) and (\ref{mnrs2}) as
well.

In the second place, we consider the case $q\in(1;\infty)$. For
$s>0$, the entropic sum is also taken as (\ref{gudf}) due to
$(1-q){\,}s<0$. Taking (\ref{lq1}) for $1\leq{q}\leq2$ and
(\ref{npqr}) for $2\leq{q}<\infty$, one gives
\begin{equation}
\left(\frac{\|\D_{\Phi}\|_{q}}{\|\D_{\Phi}\|_{1}}{\>}\frac{\|\km_{\Phi}\|_{q}}{\|\km_{\Phi}\|_{1}}\right)^{\!q}\leq
\left(\frac{\|\D_{\Phi}\|_{2}}{\|\D_{\Phi}\|_{1}}{\>}\frac{\|\km_{\Phi}\|_{2}}{\|\km_{\Phi}\|_{1}}\right)^{\!2\varkappa(q-1)}
\leq
\left\{
\begin{array}{ll}
d^{\varkappa(1-q)}{\>}, & {\mathrm{for{\ }all{\ }channels}}{\>}, \\
d^{2\varkappa(1-q)}{\>}, & {\mathrm{for{\ }unital{\ }ones}}{\>}.
\end{array}
\right.
\label{nrm1p}
\end{equation}
Here, we also used the inequality (\ref{cbn0}) and the definition
(\ref{gamdf}). Raising (\ref{nrm1p}) to the power $s>0$, the
variables $x$ and $y$ range in the domain (\ref{dadf}) with
$a=d^{s\varkappa(1-q)}$ for all quantum channels and
$a=d^{2s\varkappa(1-q)}$ for unital channels. By (\ref{gmin}), the
minimum is formally posed just as (\ref{bun0m}) but with the term
$s\varkappa$ instead of $s$ in powers of $d$ in the $q$-logarithm.
As $\gamma=1$ for $(1-q){\,}s<0$, this minimum concurs with
(\ref{mnrs1}) and (\ref{mnrs2}). For $s<0$, the entropic sum is
taken as (\ref{fudf}) due to $(1-q){\,}s>0$. Raising (\ref{nrm1p})
to the power $s<0$, the variables $x$ and $y$ range in the domain
(\ref{dbdf}) with $b=d^{s\varkappa(1-q)}$ for all quantum channels
and $b=d^{2s\varkappa(1-q)}$ for unital channels. Using
(\ref{fmin}), the minimum is formally posed as (\ref{bun0p}) but
again with the term $s\varkappa$ instead of $s$ in powers of $d$
in the $q$-logarithm. As $\gamma=2$ for $(1-q){\,}s>0$, this
minimum concurs with (\ref{mnrs1}) and (\ref{mnrs2}) as well.
$\blacksquare$

The inequalities (\ref{mnrs1}) and (\ref{mnrs2}) provide
unified-entropy trade-off relations for a given quantum channel
$\Phi$. They are an extension of the results of the paper
\cite{rprz12} with use of the family of $(q,s)$-entropies. In more
details, the physical sense of such bounds was considered in
\cite{rprz12}. When $s=1$, the formulas (\ref{mnrs1}) and
(\ref{mnrs2}) give trade-off relations in terms of Tsallis'
entropies, namely
\begin{equation}
\mathrm{M}_{q}^{(1)}(\Phi)+\mathrm{R}_{q}^{(1)}(\Phi)\geq
\left\{
\begin{array}{ll}
\gamma\ln_{q}\bigl(d^{\varkappa/\gamma}\bigr){\>}, & {\mathrm{for{\ }all{\ }channels}}{\>}, \\
\gamma\ln_{q}\bigl(d^{2\varkappa/\gamma}\bigr){\>}, & {\mathrm{for{\ }unital{\ }ones}}{\>}.
\end{array}
\right.
\label{tsrs1}
\end{equation}
Taking the limit $s\to0$, we recover the R\'{e}nyi formulation,
which is written as
\begin{equation}
\mathrm{M}_{q}^{(0)}(\Phi)+\mathrm{R}_{q}^{(0)}(\Phi)\geq
\left\{
\begin{array}{ll}
\varkappa\ln{d}{\>}, & {\mathrm{for{\ }all{\ }channels}}{\>}, \\
2\varkappa\ln{d}{\>}, & {\mathrm{for{\ }unital{\ }ones}}{\>}.
\end{array}
\right.
\label{rsrso}
\end{equation}
These trade-off relations for a single quantum channel were
originally formulated in \cite{rprz12}. Of course, we can obtain
(\ref{rsrso}) directly by taking the logarithm of both
(\ref{nrm0p}) and (\ref{nrm1p}). Since the R\'{e}nyi case was
explicitly examined in \cite{rprz12}, we refrain from presenting
the details here.

As is shown in \cite{hey06}, values of the $(q,s)$-entropy of
finite-dimensional density matrix $\bro$ are bounded from above.
Adapting this result for the considered case, we obtain
\begin{eqnarray}
\rmm(\Phi)&\leq&\frac{1}{s}{\>}\ln_{q}\Bigl(\mathrm{rank}\bigl(\D_{\Phi}\bigr)^{s}\Bigr)
\leq\frac{1}{s}{\>}\ln_{q}\bigl(d^{2s}\bigr)
\qquad (s\neq0) \ , \label{upbmn}\\
\mathrm{M}_{q}^{(0)}(\Phi)&\leq&\ln\Bigl(\mathrm{rank}\bigl(\D_{\Phi}\bigr)\Bigr)
\leq2\ln{d}
\ , \label{upbmn0}
\end{eqnarray}
since $\mathrm{rank}(\D_{\Phi})\leq{d}^{2}$. The right-hand sides
of (\ref{upbmn}) and (\ref{upbmn0}) give the upper bound on
the receiver $(q,s)$-entropy as well. The value (\ref{upbmn}) is also
obtained from (\ref{mnrs2}) for $q\in(0;1)$ and $s<0$ as well for
$q\in(1;2]$ and $s>0$. Here, the sum of the map and receiver
$(q,s)$-entropies is not less than the maximal possible value for
each of them. Thus, they may be regarded as mutually complementary
characteristics of a single quantum channel.

\section{Conclusions}\label{sec4}

We have obtained two-parametric family of entropic bounds for a
single quantum channel. These entropic relations are a
unified-entropy extension of the relations recently formulated in
\cite{rprz12}. In general, an existence of such lower bounds has
the physical relevance as some trade-off property in a single
measuring process \cite{rprz12}. For a given quantum channel, two
entropic characteristics are expressed in terms of singular values
of the dynamical and superoperator matrices, respectively. These
matrices are related to the Jamio{\l}kowski--Choi representation
and the natural representation of the channel. It is significant
that the two matrices are connected by the reshuflling operation.
Hence, they have the same Schatten $2$-norm. Entropic trade-off
relations for a single quantum channel have been formulated in
terms of the map and receiver $(q,s)$-entropies for all considered
values of the parameters. The derivation is essentially based on
some inequalities for the Schatten norms and anti-norms. Such
inequalities could be interesting in other contexts, in which we
deal with matrices related by reordering the same matrix entries.
The cases of arbitrary and unital quantum channels were both
examined. In the latter case, when bistochastic maps are treated,
we see the following. For a wide range of the parameters $q$ and
$s$, the lower bound on the sum of two entropies coincides with
the upper bound on each of the summands. This result gives an
additional reason for the fact that the map and receiver entropies
characterize mutually complementary properties of a given quantum
channel. Hence, it is interesting to study entropic bounds for
more specialized types of quantum channels, including
unistochastic maps and degradable channels. This issue could be
the subject of a separate research.

\end{document}